\documentclass[twocolumn, prd, superscriptaddress,nofootinbib]{revtex4-1}

\usepackage{array,mathtools,booktabs}
\newcolumntype{C}{>{$}c<{$}}
\AtBeginDocument{
\heavyrulewidth=.08em
\lightrulewidth=.05em
\cmidrulewidth=.03em
\belowrulesep=.65ex
\belowbottomsep=0pt
\aboverulesep=.4ex
\abovetopsep=0pt
\cmidrulesep=\doublerulesep
\cmidrulekern=.5em
\defaultaddspace=.5em
}

\usepackage{hyperref}
\usepackage[utf8]{inputenc}
\usepackage{xcolor}
\usepackage{amsmath,amssymb}

\usepackage{tikz}
\usepackage{tkz-euclide}
\usetkzobj{all}
\usetikzlibrary{decorations.pathmorphing}	% To create gluons and vector bosons
\tikzset{
    v/.style={decorate, decoration={snake, segment length=3mm, amplitude=0.75mm}, draw},
    f/.style={draw=black, postaction={decorate},
        decoration={markings,mark=at position .6 with {\arrow[very thick]{latex}}}},
    fb/.style={draw=black, postaction={decorate},
        decoration={markings,mark=at position .4 with {\arrowreversed[very thick]{latex}}}},
    fnar/.style={draw=black},
    g/.style={decorate, draw=black,
        decoration={coil,amplitude=3pt, segment length=3.5pt}},
    s/.style={dashed,draw=black, postaction={decorate},
        decoration={markings,mark=at position .55 with {\arrow[very thick]{latex}}}},
    sb/.style={dashed,draw=black, postaction={decorate},
        decoration={markings,mark=at position .55 with {\arrowreversed[draw=black,very thick]{latex}}}},
    snar/.style={dashed,draw=black,line width =1.25pt},
}

\definecolor{c1}{rgb}{0.368417, 0.506779, 0.709798}
\definecolor{c2}{rgb}{0.880722, 0.611041, 0.142051}
\definecolor{c3}{rgb}{0.560181, 0.691569, 0.194885}
\definecolor{c4}{rgb}{0.922526, 0.385626, 0.209179}
\definecolor{c5}{rgb}{0.528488, 0.470624, 0.701351}
\definecolor{c6}{rgb}{0.772079, 0.431554, 0.102387}
\definecolor{c7}{rgb}{0.363898, 0.618501, 0.782349}
\definecolor{c8}{rgb}{1, 0.75, 0}
\definecolor{c9}{rgb}{0.647624, 0.37816, 0.614037}

\newcommand{\OO}{\mathcal{O}}

\newcommand{\GeV}{\text{GeV}}

\newcommand{\eV}{\text{eV}}

\newcommand{\Mpl}{M_{\text{pl}}}
\newcommand{\Teq}{T_{\text{eq}}}

\newcounter{qnumber}

\newcounter{qnumber2}

\newcounter{qnumber3}

\def\r{\right)}
\def\l{\left(}

\def \mX {m_X}

\begin{document}

\title{Parametric Resonance Production of Ultralight Vector Dark Matter}

\author{Jeff A. Dror}
\affiliation{Berkeley Center for Theoretical Physics, Department of Physics,
University of California, Berkeley, CA 94720, USA}
\affiliation{Theoretical Physics Group, Lawrence Berkeley National Laboratory, Berkeley, CA 94720, USA}

\author{Keisuke Harigaya}
\affiliation{School of Natural Sciences, Institute for Advanced Study, Princeton, NJ, 08540}
\affiliation{Berkeley Center for Theoretical Physics, Department of Physics,
University of California, Berkeley, CA 94720, USA}
\affiliation{Theoretical Physics Group, Lawrence Berkeley National Laboratory, Berkeley, CA 94720, USA}

\author{Vijay Narayan}
\affiliation{Berkeley Center for Theoretical Physics, Department of Physics,
University of California, Berkeley, CA 94720, USA}

\begin{abstract}
Vector bosons heavier than $10^{-22}~{\rm eV} $ can be viable dark matter candidates with distinctive experimental signatures. 
Ultralight dark matter generally requires a non-thermal origin to achieve the observed density, while still behaving like a pressure-less fluid at late times.
We show that such a production mechanism naturally occurs for vectors whose mass originates from a dark Higgs. 
If the dark Higgs has a large field value after inflation, the energy in the Higgs field can be efficiently transferred to vectors through parametric resonance. 
Computing the resulting abundance and spectra requires careful treatment of the transverse and longitudinal components, whose dynamics are governed by distinct equations of motion. We study these in detail and find that the mass of the vector may be as low as $10 ^{ - 18 }~{\rm eV}$, while making up the dominant dark matter abundance. This opens up a wide mass range of vector dark matter as cosmologically viable, further motivating their experimental search.  
\end{abstract}
\maketitle

\section{Introduction}
\label{sec:Introduction}

The existence of dark matter (DM) is one of the observational evidences for physics beyond the Standard Model (SM). 
Recently, vectors ($ X _\mu $) have gained significant attention as an intriguing DM candidate with unique experimental signatures \cite{Pospelov:2008jk, Nelson:2011sf,Arias:2012az}. 
Theoretically, light vectors arise as gauge bosons of dark U(1)s, a simple extension of the SM and a common prediction of high energy theories. 
The origin of the vector mass is model-dependent and can either be a fundamental parameter in the full theory via the Stueckelberg mechanism, or can be generated through its coupling to an additional field which spontaneously breaks the corresponding U(1) via the Higgs mechanism.
In either scenario the mass of the vector is stable under quantum corrections, motivating the possibility of vectors with ultralight masses, $ m _X \ll {\rm MeV} $, limited only to having wavelengths small enough to form galaxies, $ m _X \gtrsim 10 ^{ - 22}~{\rm eV} $.

Experimentally, light relic vectors present different opportunities depending on their coupling to the SM. 
The overarching challenge in experimental prospects is competing with the powerful limits from stellar cooling~\cite{Arias:2012az,An:2013yfc} and fifth forces~\cite{Murata:2014nra} while restricting considerations to the (approximately) conserved currents of the SM (otherwise one generically expects dominant constraints from flavor changing neutral currents~\cite{Dror:2017ehi,Dror:2017nsg}). 
Nevertheless there exist many experimental proposals to search for vector DM in unexplored parameter space. 
Such states can be observed through their coupling to electrically charged particles that could be searched for in resonant cavities~\cite{Wagner:2010mi}, LC circuits~\cite{Chaudhuri:2014dla}, dish antennas~\cite{Knirck:2018ojz}, absorption in direct detection experiments~\cite{An:2014twa, Bloch:2016sjj,Aguilar-Arevalo:2016zop}, and low-energy threshold detectors~\cite{Hochberg:2016ajh, Bunting:2017net, Hochberg:2017wce, Knapen:2017ekk, Baryakhtar:2018doz}. 
If the vector couples to an unscreened force such as $ B - L $ then its coupling to neutral matter can be searched for in torsion balances and atom interferometry~\cite{Graham:2015ifn}, gravitational wave detectors~\cite{Graham:2015ifn,Pierce:2018xmy}, and pulsar binary systems~\cite{LopezNacir:2018epg}. 
With current and proposed experiments, searches for vector DM can be undertaken over almost the entire mass range $ 10 ^{ - 22}~{\rm eV} \lesssim \mX \lesssim {\rm MeV}$.

While ideas to detect vector DM are plentiful, the theoretical prospects for producing ultralight vector DM are much less explored.
%The underlying challenge is the difficulty of producing such light states by matter-radiation equality while being consistent with the requirement of cold DM. 
For light vectors there are three classes of production which have been studied in the literature: freeze-in~\cite{Redondo:2008ec}, misalignment~\cite{Arias:2012az}, and inflationary fluctuations~\cite{Graham:2015rva}. 
Freeze-in production is generically constrained by the bound on warm DM. 
Particles ``frozen-in'' through an interaction with the SM are produced with energy/momentum $ \sim T $, the temperature of the thermal bath. 
Without additional dynamics the momentum of the relics will redshift with the expansion of the universe and hence track the SM photon temperature, limiting the produced DM mass to be above a keV to be consistent with the observation of small scale structure. 
 
Misalignment has long been a standard non-thermal production mechanism for light bosons, first proposed for axions~\cite{Preskill:1982cy,Abbott:1982af,Dine:1982ah}, and later considered for massive vectors~\cite{Nelson:2011sf,Arias:2012az}. 
Here, a zero-momentum condensate of particles is produced as a result of the coherent oscillations of the field initially displaced from its minimum. 
For a generic scalar $\phi$ the energy density in the field $\rho_\phi \sim m_\phi^2 \phi^2$ remains constant when the Hubble scale is greater than its mass $H \gg m_\phi$ and the field value is stuck. 
Crucially, this is not the case for a massive vector $X$: the energy density in the vector field continues to red-shift as $\rho_X \sim m_X^2 X_\mu X^\mu \propto a^{-2}$ when $ H \gg m _X$ due to the scale factor dependence in the FRW metric on the vector norm. 
Thus any initial energy density in a massive vector field is exponentially diluted during a period of inflation, and the minimal misalignment production of vector DM fails.
This problem is avoided if an $ {\cal O} (1) $ non-minimal coupling to gravity is added to make the vector conformally invariant and hence impervious to the expansion of the universe~\cite{Arias:2012az}. 
However such a special coupling quantum-mechanically destabilizes the mass of $ X $, thus destroying one of the primary motivations for considering vector DM.

Alternatively, vector DM can be produced by the quantum fluctuations during inflation~\cite{Graham:2015rva}. 
This is a very interesting possibility, as such a production has no dangerous large-scale isocurvature perturbations and appears unique to vectors. 
Here, the observed DM abundance is saturated for $\mX \simeq 10^{-5} ~\eV (10^{14}~\GeV/H_\text{inf})^4$, and thus observational constraints on the Hubble scale during inflation~\cite{Ade:2015lrj} limit the production to masses greater than about $10^{-5} ~\eV$. 

In this paper, we propose a new production mechanism for vector DM that occurs naturally if it obtains mass through a dark Higgs field.
Generically, the production relies on a scalar field being displaced far from its minimum by the end of inflation. 
As the field rolls down its potential and begins to oscillate, its coupling to a vector results in a rapidly oscillating mass for the vector. 
This leads to non-perturbative production of $X$ particles through a parametric resonance (PR) instability (as is the case in theories of reheating~\cite{Dolgov:1989us, Traschen:1990sw, Kofman:1994rk,Kofman:1997yn}, though the dynamics we consider take place solely during radiation domination). 
Crucially, the rate of production is much greater than that of any possible perturbative process. 
The produced particles then have more time to red-shift, significantly relaxing the coldness constraint and allowing for the production of ultralight DM. 
This is in analogy with earlier work~\cite{Co:2017mop} on the PR production of axion DM via dynamics of a Peccei-Quinn symmetry breaking field (for other work on non-perturbative production of relics, see~\cite{Felder:1998vq,Mazumdar:2015pta,Agrawal:2017eqm,Kitajima:2017peg}).
In this paper we focus on the minimal case where the scalar field is a dark Higgs.
The nature of the resonance and resulting abundance of vectors and dark Higgses is different depending on the strengths of the gauge coupling $e$ and dark Higgs quartic coupling, $\lambda  $.
We examine both limits and find that vector DM can be produced with masses as light as $m _X  \gtrsim  10^{-18} ~{\rm eV}$, consistent with all constraints. 
This opens up most of the mass range for vector DM as cosmologically viable, and further motivates the experimental program searching for such particles. 

The paper is organized as follows.
In Sec.~\ref{sec:Model} we outline the model of interest and show the limitations of a perturbative Higgs decay in producing light vector DM. 
In Sec.~\ref{sec:PR} we review the relevant non-perturbative dynamics of PR, specifically as it applies to vector production. 
In Sec.~\ref{sec:vectorDM} we examine the PR production of ultralight vector DM, and in Sec.~\ref{sec:pheno} we discuss additional cosmological consequences and constraints on the mechanism.
Finally, in Sec.~\ref{sec:discussion} we conclude and discuss future directions.

\section{The Model}
\label{sec:Model}
We now present an outline of the model and detail the dynamics of an oscillating scalar field in the potential. 
As our starting point we consider a complex scalar field, $\varphi$, that will give a mass to the vector:
\begin{equation}
\label{eq:modelL}
{\cal L}  = -\frac{1}{4} X_{\mu \nu}X ^{\mu\nu}  + |D_\mu \varphi|^2 - V(\varphi)\,,
\end{equation}
where $D_\mu = \partial_\mu + i e X_\mu$, $e$ is the dark gauge coupling constant (we absorb the scalar charge into the definition of $e$), and $ X _{\mu\nu} $ is the field strength tensor. We consider the simplest model of spontaneous symmetry breaking with a potential parameterized as 
\begin{equation}
\label{eq:Vphi}
V(\varphi) = \lambda^ 2 \left( \left|  \varphi \right| ^2-\frac{v^2}{2} \right) ^2\,.
\end{equation}
Expanding $ \varphi $ around the vacuum expectation value (VEV), we obtain:
\begin{equation}
\label{eq:modelL2}
{\cal L}  \supset \frac{1}{2} e ^2 v ^2  \left( 1 + \frac{ \phi }{ v } \right) ^2X_\mu X^\mu   - V(\phi)\,,
\end{equation}
where
\begin{equation}
\label{eq:Vphi2}
V(\phi) = \frac{1}{4} \lambda ^2  \phi ^2 \left( \phi + 2 v \right) ^2\,.
\end{equation}
The vacuum masses of $ X $ and dark Higgs boson are $m _X  = e v$ and $m _\phi  = \sqrt{2} \lambda v$, respectively. \footnote{Note that while the vector mass is radiatively stable, the scalar mass is not and naturalness would suggest a cut-off of order $ \Lambda \lesssim v \min\{1, m_\phi/m_X\}$.
Ultimately we will be interested in VEVs much larger than the weak scale, so fine-tuning in the dark Higgs sector is not a serious constraint and we will not address it further. }
Furthermore, we refrain from making any assumptions about the magnitude of the vector coupling to the SM, up to assuming the coupling is not so large that it efficiently thermalizes the two sectors (or is phenomenologically excluded in other ways). 

We assume $\phi$ starts out displaced from its minimum after inflation with an initial field value, $ \phi _0 $.  
The classical equation of motion for $\phi$ is
\begin{equation} 
\label{eq:phiEOM}
\ddot{\phi}
+  3 H \dot{\phi} + \lambda ^2 ( \phi ^3 + 3 v \phi ^2 + 2 v ^2  \phi ) = 0\,,
\end{equation} 
which is valid as long as the back-reaction due to any created particles is negligible (these effects are crucial in the termination of non-perturbative particle production and will be addressed later).
The field is stuck until $H \sim m_{\rm{eff}}(\phi_0)$ where $m_{\rm{eff}}(\phi) = \sqrt{V''(\phi)}$ is the effective (field-dependent) mass, at which point $\phi$ begins oscillating about the minimum. As long as $ \phi _0 \ll M_{\rm pl} $ (regardless of the hierarchy between $ \phi _0 $ and $ v $) the universe is radiation-dominated at the onset of oscillations which begin at,
\begin{equation}
\label{eq:Tosc}
T_\text{osc} \simeq 0.5 \sqrt{ m_{ {\rm eff}} M_{\rm pl} }\,,
\end{equation}
where $M_{\rm pl}  = 2.4 \times 10^{18} ~{\rm GeV} $ is the reduced Planck mass. 

We now consider the two limits for the initial field value, $ \phi_0 \ll v $ and $ \phi_0 \gg v $.
If $ \phi_0 \ll v $, oscillations start at $ T _{ {\rm osc}}$ and the solution is the well-known harmonic oscillations, $\phi ( t )  = \Phi \cos  \left( m _\phi t \right)$.
The amplitude of oscillations red-shifts with the scale factor $ a $ (we use the convention that $a=1$ at the onset of oscillation) as $\Phi(t) = \phi_0 a^{-3/2}$, and the energy density in coherent oscillations acts as non-relativistic matter $\rho_\phi \propto a^{-3}$. 

Conversely, if the field value is large, $ \phi_0 \gg v $, then the effective mass is $m_{\rm{eff}}(\phi_0) \simeq  \sqrt{3} \lambda \phi_0 $ with $ T _{ {\rm osc}} \simeq   \sqrt{\lambda \phi_0 M_{\rm pl} }$.
Due to the conformal invariance of the quartic potential, it is most convenient to switch to conformal coordinates. Furthermore, it is convenient to absorb the oscillation time into our definition suggesting the coordinate transformation $ \bar{\phi} \equiv a \phi / \phi _0  $ and $ d z \equiv \lambda \phi _0 d t /  a $.
The equation of motion is then simply:
\begin{equation}
\bar{\phi}  '' +  \bar{\phi} ^3 = 0\,,
\label{eq:phiconf}
\end{equation}
where we use primes to denote derivatives with respect to $ z $. The exact solution is an elliptic cosine function with elliptic modulus of $ 1/2  $, 
\begin{equation} 
\bar{\phi} ( z ) =  {\rm cn} (z)\,.
\label{eq:elliptic}
\end{equation} 
This function is usually well-approximated by the simple cosine function, $\bar{\phi} \simeq  \cos(0.85 z) $, the first term in its Lambert series expansion, but some features require keeping higher order terms and so we refrain from making this approximation. Here, the (original) field amplitude instead red-shifts as $\Phi(t) = \phi_0 a^{-1}$ and the energy density in coherent oscillations acts like radiation $\rho_\phi \propto a^{-4}$. 

\subsection{Perturbative Decay}
\label{sec:pert}
The dynamics of particle production depend critically on the initial field value, and we postpone a careful treatment to Sec.~\ref{sec:PR}. However, we generally expect non-perturbative effects are negligible if $ \phi _0  \ll v $ and we briefly review the physics in this limit.
We first compute the production of vector DM from perturbative decay of the dark Higgs---this will eventually highlight the effectiveness of parametric resonance. 
Coherent oscillations of the $\phi$ field result in an yield of dark Higgs:
\begin{equation}
\label{eq:DHyield}
Y_\phi = \frac{  \rho _\phi }{ m _\phi s  } \simeq  \frac{0.5}{\lambda^{1/2}} \l\frac{\phi_0}{v}\r^{2} \l\frac{v}{\Mpl}\r^{3/2},
\end{equation}
where $s$ is the entropy density.
This population can decay into $ X $ if it is kinematically allowed, i.e., $ m _\phi > 2 m _X    $. 
Since the co-moving number density in the dark sector is conserved, the dark Higgs condensate will fully convert into a co-moving number density of vectors $Y_X = 2 Y_\phi$. 
The timescale for this conversion is set by the decay rate $\Gamma_{\phi \to X X}$, which is dominated by the decay into longitudinal modes of $X$: 
\begin{equation}
\label{eq:invisdecay}
\Gamma _{  \phi \rightarrow XX} \simeq  \frac{m _\phi ^3  }{32\pi v ^2 } \,.
\end{equation}
The underlying challenge with DM production via decays is that the $X$ particles are initially highly boosted with momentum $\OO(m_\phi)$.
In this case, the produced vectors begin red-shifting as non-relativistic matter once the universe cools to a temperature
\begin{equation} 
\label{eq:DPcoldness}
T _{ {\rm NR}} \simeq 0.1~m _X \left( \frac{ M_{\rm pl} \lambda }{ v } \right) ^{1/2}.
\end{equation} 
From here on $T_\text{NR}$, and in general the term ``temperature'', refers to that of the SM thermal bath (this is distinct from a possible dark sector temperature, which may or may not even be in thermal equilibrium). 
As expected, $T_\text{NR}$ increases with $\lambda$ which corresponds to earlier decays. 
Observations of cosmological large-scale structure require that the DM be non-relativistic by around a $ {\rm keV} $ and so we require $ T _{ {\rm NR}} \gtrsim  {\rm keV} $~\cite{Lopez-Honorez:2017csg, Irsic:2017ixq,Ade:2015xua} (precise constraints range from $ \sim 1 -5~{\rm keV} $ though suffer from astrophysical uncertainties).
Based on~\eqref{eq:DHyield} and~\eqref{eq:DPcoldness}, we find the vector abundance equals the relic density of DM for masses:
\begin{equation} 
\label{eq:mXperturbative}
m _X \simeq 10 \left( \frac{ T ^3 _{ {\rm NR}} T _{ {\rm eq}} }{ \lambda } \right) ^{1/4} \l \frac{v}{\phi_0} \r^{1/2},
\end{equation} 
where $ T _{ {\rm eq}} \simeq 0.75~  {\rm eV}  $ is roughly the temperature at matter-radiation equality.
Note that production of light vector DM here favors large values of $\lambda$, which is ultimately limited by perturbativity $\lambda < 2\pi$ (this is in fact a stronger condition than $\phi_0 < M_{\rm pl} $). 
Saturating the coldness and perturbativity constraints, we conclude that it is impossible to produce vector DM with mass less than a keV using perturbative decays of the scalar field. 

\section{Parametric Resonance}
\label{sec:PR}

If $ \phi_0 \gg v $, the production rate of vectors can be much larger than the perturbative rate. Such a large initial field value is a generic expectation unless the coupling with an inflaton strongly fixes $\phi$ to the origin. 
In the classical background of an oscillating $\phi$ field, the field $ X $ feels a large, oscillating, mass.
This may lead to a period of non-perturbative, exponential production of vectors though parametric resonance (PR).
\footnote{This is distinct from tachyonic resonance, which is an exponential instability that occurs for modes with a negative effective frequency-squared. 
}
Particle production by PR is a well-studied phenomena, particularly in the context of reheating after inflation (so-called {\em preheating})~\cite{Dolgov:1989us, Traschen:1990sw, Kofman:1994rk,Kofman:1997yn}.
However, vector production by PR has not been studied nearly as extensively as for scalars. 
PR production of gauge fields at the end inflation has been previously considered in~\cite{Finelli:2000sh,Lozanov:2016pac, Ema:2016dny}, e.g. to seed primordial magnetic fields~\cite{Kandus:2010nw}. 
In addition,~\cite{Lozanov:2016pac} and~\cite{Ema:2016dny} also discuss the enhanced production of longitudinal modes.   
In this section we review the theory of PR for a vector, and show that the dynamics in our case depend delicately on the hierarchy between couplings $ e$ and $ \lambda $ and require a careful treatment of the transverse and longitudinal modes. 
We present the differential equation which governs the production of longitudinal modes that is distinct from the well-studied Mathieu and Lam\'{e} equations (the typical differential equations studied in the context of PR). 
The classes of solutions are presented in an instability chart of the exponentially growing momentum modes as a function of $e/\lambda$, and we compare the PR production of longitudinal and transverse modes in the different limits of interest. 
Ultimately we show that, as a consequence of both initial conditions set by inflation and a longitudinal mode enhancement in the coupling, the longitudinal mode dominates production for a wide range of couplings.

\subsection{Parametric Resonance for a Higgsed Vector}
Using the conventional $ {\rm diag} (1,- a ^2 , - a ^2 , - a ^2 ) $ metric in an expanding universe, we can write the kinetic and mass term of $ X $ explicitly in terms of its temporal and spatial components:
\begin{align} 
\frac{1}{4} X _{\mu\nu} X ^{\mu\nu} & = \frac{1}{2} \left[ \frac{1}{ a ^2 } \left| \partial _t {\bf X} + \nabla X _t \right| ^2 - \frac{1}{ a ^4 } \left| \nabla \times {\bf X} \right| ^2 \right] \,,\\
\frac{1}{2} \tilde{m}   ^2 X _\mu X ^\mu & = \frac{1}{2}\tilde{m}  ^2 \left( X _t ^2  - \frac{1}{ a ^2 } \left| {\bf X} \right| ^2 \right) \,,
\end{align} 
where $ \tilde{m} \equiv m _X  \left( 1 + \phi  / v \right)$. 
Since $ X _t $ does not contain a kinetic term, it is an auxiliary field and can explicitly be integrated out using its equation of motion. 
Switching to $k$-space, we separate $\bf{X}$ into its longitudinal and transverse components such that $ k \cdot {\bf X} = k X_L $ and $ {\mathbf{k}} \cdot {\bf X} _T = 0 $. 
As a result, the action for the vector field separates for the transverse and longitudinal components $S = S_T + S_L$:
\begin{align} 
S _T & = \int dt  \frac{a^3 d^3k }{ (2\pi)^3 } \frac{1}{2 a^2} \left( \big| \dot{ {\bf X}} _T \big| ^2 - \left(  k ^2/ a ^2  + \tilde{m}  ^2 \right)  \big|{\bf X} _T \big| ^2 \right) \,, \\ 
 S _L & =   \int dt  \frac{a^3 d^3k }{ (2\pi)^3 } \frac{1}{2 a^2} \left( \frac{ \tilde{m} ^2 a ^2  }{  k ^2 + \tilde{m} ^2 a ^2  }  \dot{  X} _L  ^2 - \tilde{m} ^2   X _L ^2  \right) \,.\label{eq:slong}
\end{align} 
Note for here and throughout we use $ k $ to denote co-moving momentum.

\begin{figure} 
  \begin{center} 
\includegraphics[width=8cm]{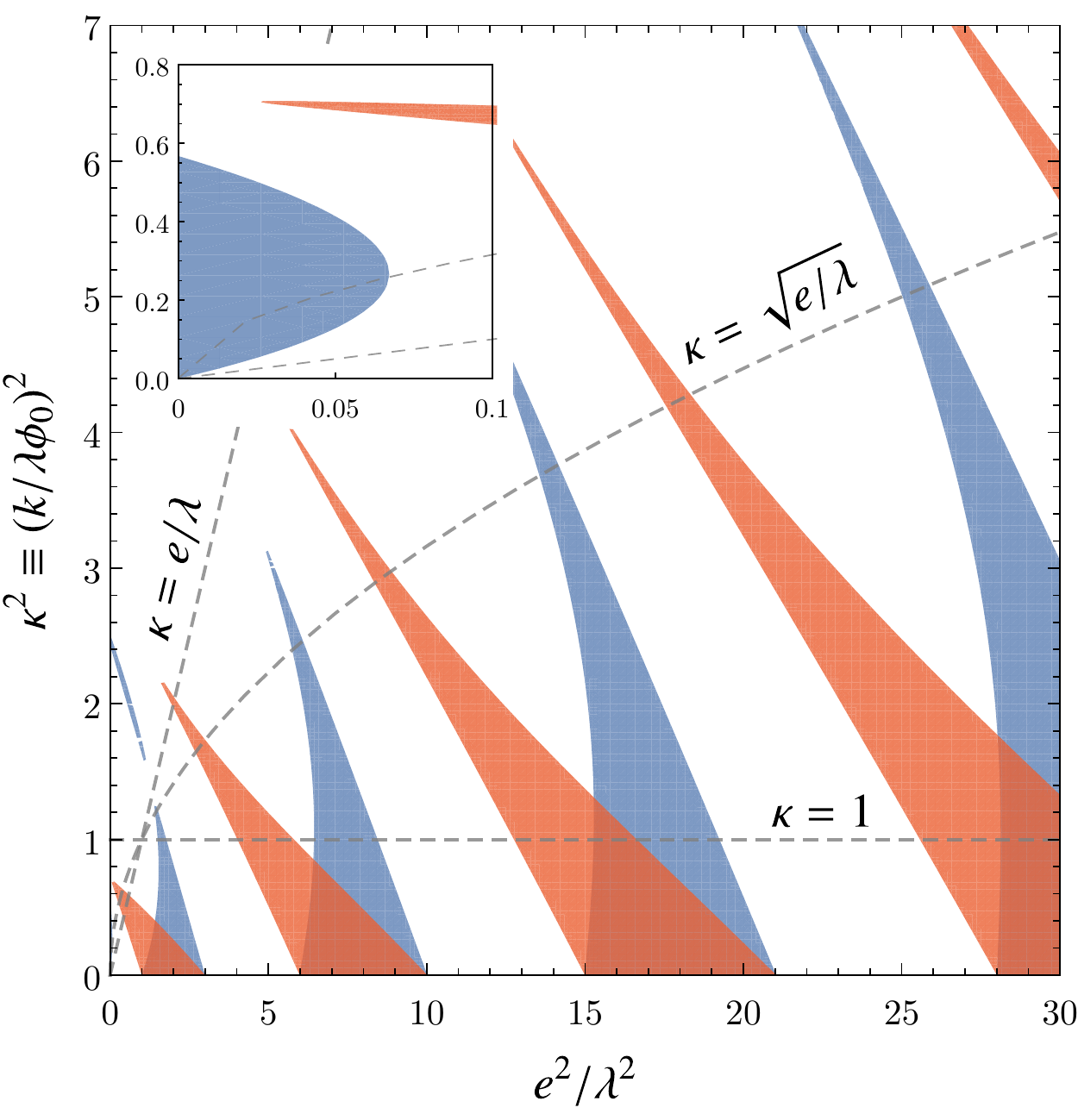} 
\end{center}
\caption{Instability charts of transverse ({\bf \color{c4} red}) and longitudinal ({\bf \color{c1} blue}) modes. The dashed lines represent different values of $ \kappa $, and resonance bands above $ \kappa = e/ \lambda $ correspond to relativistic production. In the inset we show the $ e / \lambda \ll 1$ limit in which the enhancement of longitudinal mode production over transverse modes is seen explicitly. 
} 
\label{fig:instability}
\end{figure}

We now study the PR production of the transverse and longitudinal modes. 
Exponentially growing modes naturally occur in specific resonance bands, and it is conventional to map out the regions of unstable momentum as a function of the couplings. 
The dominantly produced modes lie in the widest resonance band and generally have a large exponential instability, resulting in a rapid conversion of the energy density in the oscillating scalar field into these modes.
We present the instability charts for the transverse and longitudinal modes in Fig.~\ref{fig:instability} and refer to it throughout---this is obtained by numerically solving the relevant equations of motions and identifying the choice of couplings that result in exponentially-growing solutions. 

\paragraph{Transverse Modes}
It is convenient to switch to dimensionless, conformally invariant quantities: a time variable, $ d z \equiv \lambda \phi _0 d t / a $ as introduced in~\eqref{eq:phiconf}, a momentum $ \kappa \equiv k / \lambda \phi _0 $, and a conformal oscillating mass, $  \mu \equiv a \tilde{m} / \lambda \phi _0 = ( e / \lambda ) {\rm cn} (z) $.
Doing so, the equation of motion for transverse modes becomes:
\begin{equation}
\label{eq:Teqn}
{\bf{X}}_T'' + \left( \kappa ^2+ \mu ^2 (z)   \right) {\bf{X}}_T = 0\,. 
\end{equation}
This equation is known as the Lam\'{e} equation and has been extensively studied in the literature (see e.g.~\cite{Greene:1997fu}).~\footnote{As long as the scalar oscillations are well approximated by the harmonic approximation, the solutions are the same as those of the well-known Mathieu equation.}
Solutions to this equation are exponentially growing for certain momentum modes ${\bf X}  _T  \propto e^{\mu_\kappa  z}$. 
The characteristic exponents, $ \mu _\kappa  $, are a non-trivial function of momentum as well as the ratio of couplings. 
PR is often classified as either broad or narrow, based on the width of resonance bands and the size of the characteristic exponents. 
For the mode equation of~\eqref{eq:Teqn}, the resonance is broad if $e \gg \lambda$ and narrow if $ \lambda \gg e $. 
We will be interested in both these limits, which have previously been solved analytically. 

In the case of $ \lambda \gg e $ the first resonance band around $ \kappa ^2  \simeq 1 $ dominates production while subsequent resonance bands (at larger $\kappa$) become increasingly narrow.
Such narrow resonances are known to have suppressed production with a small range of produced momenta and characteristic exponents, $ \Delta \kappa^2 , \mu_\kappa \propto e ^2 / \lambda ^2 $.
Thus we conclude production of transverse modes is not efficient in this regime.
For $e \gg \lambda$ the resonance is instead broad and can achieve much more efficient production. 
Inspection of Fig.~\ref{fig:instability} shows that the structure of the resonance, in particular the size of $\kappa$ in the first resonance band, depends critically on the value of $e/\lambda$. 
Interestingly, it is still the case that for $ e \gg \lambda $ there is an upper bound on the produced momentum which can be estimated analytically~\cite{Greene:1997fu}. 
A necessary condition for exponential instability in the regime of broad PR is non-adiabatic change in the frequency of fluctuations. 
The (dimensionless) frequency felt by the transverse modes is $\omega(t) \simeq  \sqrt{\kappa ^2 + ( e / \lambda ) ^2   {\rm cn} ^2  (z)}$. 
If one defines an adiabatic parameter $ R \equiv |\omega'|/ \omega^2 $, then for most of the oscillation period this is close to zero and the frequency changes adiabatically.
The only time $ R > 1 $ is when the background field oscillates toward zero, ${\rm cn} (z)  \to 0$ and $ \kappa ^2  \lesssim e / \lambda $, which is an estimate of the upper bound on the dominantly produced momenta. 
In fact this is bound is evident in Fig.~\ref{fig:instability}, where the widest resonance band (red) always lies below the line $\kappa =  \sqrt{ e / \lambda } $.
We thus find that the typical physical momenta produced by PR here is much less than the time-averaged mass of the vector, $ \sim e \phi _0 $, and vectors are produced non-relativistically. 
We compute the maximum characteristic exponent numerically for $e \gg \lambda$ and find $\mu_\kappa \simeq 0.2$, in agreement with the previous literature~\cite{Greene:1997fu}.

\paragraph{Longitudinal Mode}

PR for the transverse modes reduce to equations that have been solved extensively in the literature.
We now move to the longitudinal mode which, as we will show, dominates the production of vectors in a wide range of parameters. 
Starting from~\eqref{eq:slong} and making the transformations to conformal fields, we find the equation of motion:
\begin{equation}
\label{eq:Leqn}
X'' _{ L} + \frac{ 2 \kappa  ^2  }{ \kappa  ^2 + \mu  ^2  } \frac{ \mu ' }{ \mu }  X' _{ L}   + \left(  \kappa  ^2  + \mu ^2  \right)  X _{ L} = 0\,.
\end{equation}
The dynamics governed by this differential equation have not been studied in great detail in the literature. 
Here we present a brief analysis, and leave an extensive study for future work. 

First we note that in the limit $\kappa \to 0$, the equation of motion~\eqref{eq:Leqn} reduces to precisely that of the transverse modes:
\begin{equation}
 X_L'' + (   \kappa^2 +\mu^2)  X_L \simeq 0\,.
\end{equation}
This is expected, since at low energies the longitudinal mode can no longer be distinguished from the transverse modes and should obey the same dynamics. 
We also see this directly in Fig.~\ref{fig:instability}, where the resonance bands of the two modes roughly coincide (except for very particular values of the couplings) in the limit of small momentum $ \kappa \ll 1$.

The high energy limit is more challenging to analyze since the physics is obscured by a divergence in the friction term as the oscillating field passes through the origin. 
While it is in principle possible to solve the equation as is, it is simpler to introduce a field redefinition,
\begin{equation} 
\pi \equiv \frac{ \mu }{ \kappa } X _L\,.
\end{equation}
Being a linear transformation, this does not mix the different momentum modes and hence does not obscure the structure of the resonance.
The resulting equation of motion is:
\begin{equation}
\label{eq:GBeomfull}
\pi '' -  \frac{ 2 \mu \mu ' }{ \kappa ^2 + \mu ^2 } \pi ' + \left( - \frac{ \mu '' }{ \mu } + \frac{ 2 \mu ^{ \prime 2 } }{ \kappa ^2 + \mu ^2 } + \kappa ^2 + \mu ^2  \right) \pi = 0 \,.
\end{equation}
If we then take the high-energy limits, $\kappa \gg \mu $ and $ \mu \ll  1$, we recover a familiar form:
\begin{equation}
\label{eq:GBeom}
\pi'' +  (\kappa^2 + \text{cn}^2(z))  \pi \simeq  0\,.
\end{equation}
This is analogous to the equation of motion for transverse modes~\eqref{eq:Teqn}, though crucially the amplitude of the oscillations is enhanced by a factor $ \lambda^2/ e^2 $.
As a result, the PR dominantly produces longitudinal modes with $ \kappa^2 \simeq 1 $, which can also be seen directly in the inset of Fig.~\ref{fig:instability} where there is a wide instability band (blue) for the longitudinal mode in the limit $ e / \lambda \rightarrow 0 $. 
This result in the high-energy limit can also be derived directly from the action of the dark Higgs $ \varphi $ using the Goldstone boson equivalence theorem.
Expanding $\varphi = (\phi + v + i \chi)/ \sqrt{2} $ and switching to conformal fields, we find the same equation of motion for $\chi$ as found for the longitudinal mode in this limit~\eqref{eq:GBeom}.

We are now in a position to complete the discussion of PR for the longitudinal mode as a function of the couplings. 
Firstly we consider the limit of $\lambda \gg e$ (where we found the transverse modes are not efficiently produced). 
In this case the longitudinal mode is produced strictly in the high-energy regime $ \kappa \gg \mu $, and the results follow the approximate form of the mode equation~\eqref{eq:GBeom}. 
Here we find the resonance is efficient for $\kappa^2 \simeq \Delta \kappa^2 \simeq 1$ and we again have a large characteristic exponent $\mu_\kappa \simeq 0.1$.
We emphasize that, in contrast to the transverse modes, the longitudinal mode has a marginally narrow resonance allowing it to be produced efficiently.
In addition, the longitudinal modes are produced highly boosted with relativistic momenta. 
We now turn to the limit $ e \gg  \lambda $ which is much more interesting. 
Although the mode equations become identical in the limit $\kappa \ll 1$, PR production only occurs when the vector mass is rapidly varying (i.e. when adiabaticity is violated). 
At this point $\kappa$ is of order the oscillating mass $\mu$, and the longitudinal mode equation~\eqref{eq:Leqn} does not approximately reduce to any well-known forms (due to the non-negligible friction term). 
Indeed, as is evident from Fig.~\ref{fig:instability}, there are substantial differences between the resonance structures of the longitudinal and transverse modes in this regime. 
While the solutions for the longitudinal mode similarly suggest an upper bound on the dominantly produced momenta, the bound may be larger than that of the transverse modes depending on the coupling. 
This is an intriguing feature that opens up the possibility of producing relatively boosted longitudinal modes, although we still expect that the momenta in the first resonance bands satisfy $\kappa^2 < e/\lambda$ such that produced modes are not relativistic. 
Finally, we compute the typical characteristic exponent for longitudinal mode production in this limit to be $\mu_\kappa \simeq 0.2$.

\paragraph{$ \phi $ Fluctuations}
In addition to vector production, an oscillating $\phi$ field will inevitably also resonantly produce $\phi$ fluctuations with non-zero momentum (denoted as $\delta \phi$ to differentiate from the zero-momentum condensate which we continue to denote by $ \phi $) from the self-coupling, $\lambda$. 
We emphasize that these excitations are in addition to the zero-mode condensate that results from coherent oscillations and carry a particle interpretation similar to the vector fluctuations. 
The mode equation can be derived from~\eqref{eq:phiEOM} by restoring the momentum term and expanding the field as $ \phi  + \delta \phi $, keeping order linear terms in the fluctuations. 
The resulting equation of motion is identical to that of the transverse modes~\eqref{eq:Teqn} but with the replacement $e^2 \to 3 \lambda^2$:
\begin{equation}
\label{eq:modephi}
\delta \phi'' + \left(  \kappa^2 + 3 {\rm cn}   ( z )^2 \right)   \delta \phi  = 0\,.
\end{equation}
The PR is qualitatively similar to that of the longitudinal mode in the $ \lambda \gg e $ case~\eqref{eq:GBeom}. 
Fluctuations of $\phi$ are dominantly produced at momentum $\kappa^2 \simeq 1$ with a width $\Delta \kappa^2 \simeq 1$ and (slightly smaller) characteristic exponent $\mu_\kappa \simeq 10^{-2}$. 

\paragraph{Initial Conditions}
We have seen that due to parametric resonance, there is an exponential amplification of fluctuations in the fields $X$ and $\phi$ for certain momentum modes. However, an important effect we have yet to address are the initial conditions for the fields. Assuming a period of inflation, we can estimate the initial conditions for each field. Transverse components of the vector are conformally invariant and do not experience the expansion suggesting that they should have an initial field value given by the Bunch-Davies vacuum with a power spectrum, $ P _T  ( k ) \sim k ^2 $. The initial conditions of the longitudinal mode are more dramatic. These are created by coupling to the metric during inflation and can far exceed their transverse counterparts~\cite{Graham:2015rva} with a power spectrum, $ P _L  ( k ) \sim ( H _\text{inf} k / e \phi _0 ) ^2 $ (this applies for both $ \lambda \gg e $ and $ e \gg \lambda $ and assumes the vector mass during inflation is $ e \phi _0 $). This gives a ratio of the longitudinal to transverse mode amplitudes at the end of inflation as,
\begin{equation} 
\frac{ X _{ L } (k) }{ X _{ T} (k) } \gtrsim  \frac{  H _{ {\rm inf}} }{ e \phi _0 }\,,
\end{equation} 
which is independent of $  k $. Since we do not consider parameter space such that the vector mass is above the scale of inflation, the longitudinal mode will dominate the transverse mode production as long as they can both be produced efficiently. The scalar fluctuations during inflation behave similarly to the longitudinal mode and will have comparable initial conditions.

\subsection{Final Relic Abundance and Momenta}

The exponential production from PR does not last indefinitely. 
Thus far we have neglected the non-linear back reaction of these fluctuations on PR itself. 
There are three kinds of back reactions:

(1) The vector and scalar fluctuations grow large enough and give large mass contributions to both $ \phi $ and $ X $ that subsequently red-shift as $\propto 1 / a $, and can lead to other interesting cosmological effects that will be discussed later. 
Here, we see that a changing mass acts to shift the resonance bands and can thus ruin the important Bose enhancement in final states that leads to continued exponential production for growing modes.

(2) Scattering of fluctuations with the zero-mode condensate as well as fluctuations shift the particle momenta out of resonance bands.  Again, this destroys the Bose enhancement in produced fluctuations and can also shut down exponential production. 

(3) The scattering also depletes the zero-mode condensate and terminates PR.

A fourth effect, due to the expansion, is not present in this theory due to its conformal nature.
In practice these effects occur simultaneously and act to cease particle production when the energy density of the fluctuations becomes comparable to the original energy density in the condensate. 
While these effects are highly non-linear and challenging to compute, if particle production lasts long enough the condensate will completely convert into the produced $ \phi $ and $ X $ particles, regardless of the detailed processes involved. We assume that the zero-momentum field is completely depleted and does not make up any of the DM today (we expect this is a reasonable approximation due to significant scattering with produced fluctuations at the end of PR). 
In this sense, a full solution to the equations of motion, including back reactions, gives us the relative fraction in these two populations. 
We can parameterize the yields after the conclusion of PR production as:
\begin{equation} 
 Y _{ X } = f \frac{\rho _{ \phi , {\rm osc}} }{E_X s(T_\text{osc})}
 \,, \,Y _{ \delta \phi } = (1-f) \frac{\rho _{ \phi , {\rm osc}} }{E_{\delta \phi} s(T_\text{osc})}\,.
\label{eq:ndensity}
\end{equation}  
Here $ \rho _{ \phi , {\rm osc}} = \frac{1}{4} \lambda ^2 \phi _0 ^4$, $ f $ is the relative fraction of the condensate co-moving energy density dumped into vectors, and $ E _i ^2  =  \sqrt{ k _{\ast ,i} ^2 +  m _i    ^2     } $ are the co-moving energies of the particle species $i$ ($ m _i $ denotes the time-varying mass of the particle).
For simplicity, we assume particles are produced with a co-moving momenta peaked at $k_{\ast, i}$ though in practice there will be small corrections associated with an ${\cal O}(1)$ spread around this typical value.

Once PR stops being efficient the produced particles are in a highly non-equilibrium state with peaked momenta. These particles can still undergo collisions within the sector scattering their momenta and changing their number densities. This includes simple $ 2 \rightarrow 2 $ elastic scattering as well inelastic processes inducing cannibalization.
Tracking the dynamics rigorously throughout this ``post-scattering'' phase requires a dedicated computation putting vectors and scalars on lattices and is beyond the scope of this work. 
However, we can still qualitatively estimate the behavior in each limit.

For $  \lambda \gg e $, the longitudinal mode and the fluctuations only differ by their parity and hence have comparable energy densities after decay ($ f \simeq 1/2 $) as well as momenta $ k _\ast \sim \lambda \phi _0 $. In this limit, the symmetry between the longitudinal mode and the fluctuations of $ \phi $ allows us to treat them as a single fluid regardless of the details of the post-scattering phase. 
Furthermore, we do not expect these processes to be active even if either species becomes non-relativistic. 
We thus expect that the vectors and scalars should have comparable number densities and momenta at late times:
\begin{equation} 
Y _X , Y _{\delta \phi }\rightarrow \frac{1}{2} \frac{\rho _{ \phi , {\rm osc}} }{\lambda \phi_0 s(T_\text{osc})}\,.
\end{equation} 

For $ e \gg \lambda $, the situation is more subtle. 
As shown, vectors are primarily produced non-relativistically from PR (a detailed spectrum will depend on the coupling $ e/ \lambda $), while fluctuations of $ \phi $ are produced mildly relativistically. 
At this point, elastic and inelastic processes are efficient in driving the sector toward a state equating the momenta and number densities of $ X $ and $ \phi $ ($ f \simeq 1/2 $). 
Effective scattering during this time relies on a Bose enhancement of the final state which is spoiled at large enough momenta.
We estimate that such processes cannot produce vectors with momenta larger than their mass $ e \phi _0 $, and as a result both species should eventually be up-scattered to a co-moving momenta as large as $k_\ast \sim e \phi_0$.
We thus expect yields of $X$ and $\phi$ at late times of order:
\begin{equation} 
Y _X , Y _{\delta \phi }\rightarrow \frac{1}{2} \frac{\rho _{ \phi , {\rm osc}} }{e \phi_0 s(T_\text{osc})}\,.
\end{equation} 
We have confirmed this expectation employing a lattice computation using LATTICEEASY~\cite{Felder:2000hq} and approximating the vector interaction by that of a scalar field with a quartic coupling to $\phi$.

\section{Vector Dark Matter from Parametric Resonance}
\label{sec:vectorDM}

The above results apply for any Higgsed vector in the early universe, and we now consider the implications of PR for the production of ultralight vector DM. For the rest of this section, we assume a large initial field value $ \phi _0 \gtrsim v $.
In practice, PR production is not instantaneous but requires sufficiently long exponential growth so we in fact have the condition $ \phi _0/v \gtrsim 10-100 $.\footnote{A stable PR still occurs even if $\phi_0 \lesssim v$ in the case $  \lambda \gg e $, but this is a narrow resonance and highly inefficient.}

There are four (a priori) independent parameters in the model: $\{\phi_0, v, \mX, m _\phi \}$ or alternatively $\{\phi_0, v, e, \lambda \}$. 
As we have seen, the nature of PR depends on the relative strengths of the couplings in a non-trivial way. 
This is also true for the resulting constraints on safely obtaining the correct relic abundance of vector DM. 
Thus we look at the two simplifying limits separately: $ \lambda \gg e $ and $ e \gg \lambda $. 
Here we focus on the fundamental challenge of being consistent with constraints on warm DM while producing the entire DM abundance.
Additional constraints and phenomenological consequences of the vector production are examined in Sec.~\ref{sec:pheno}, which we refer to in the results of Fig.~\ref{fig:largelambda} and~\ref{fig:smalllambda}.

\subsection{Case 1: $ \lambda \gg e $}

We begin with the case where the gauge coupling is small with respect to the quartic (and hence also $ m _X \ll m _\phi $). 
In Sec.~\ref{sec:PR} we estimated a yield for the $X$ and $\phi$ fluctuations to be roughly equal at late times:
\begin{equation}
\label{eq:yields1}
Y_{X} \simeq Y_{\delta \phi} \simeq \frac{0.01}{\lambda^{1/2}} \left(  \frac{\phi_0}{\Mpl} \right) ^{3/2}\,.
\end{equation}
In the absence of any additional interactions these yields are conserved until today. Once $X$ and $\phi$ become non-relativistic, $X$ constitutes a small fraction of the energy density of the dark sector:
\begin{equation} 
\frac{ \Omega  _{ X } }{ \Omega  _{ {\rm DM}}  } \simeq \frac{ m _X }{ m _\phi } \sim \frac{ e }{ \lambda } \,. 
\end{equation} 
Furthermore, the typical co-moving momenta of each species is of order $k_* \sim \lambda \phi_0$. 
This is related to the physical momenta by red-shifting from the time of production.
Crucially, particles are produced from PR at very early times near the start of oscillations.
Due to the conformal invariance, we can effectively treat the yields~\eqref{eq:yields1} as being produced with a physical momenta $\lambda \phi_0$ at a temperature $T_\text{osc}$ even if the particles are dominantly created somewhat later (PR results in rapid, through not instantaneous, particle creation).

\begin{figure*}
\includegraphics[width=8.25cm]{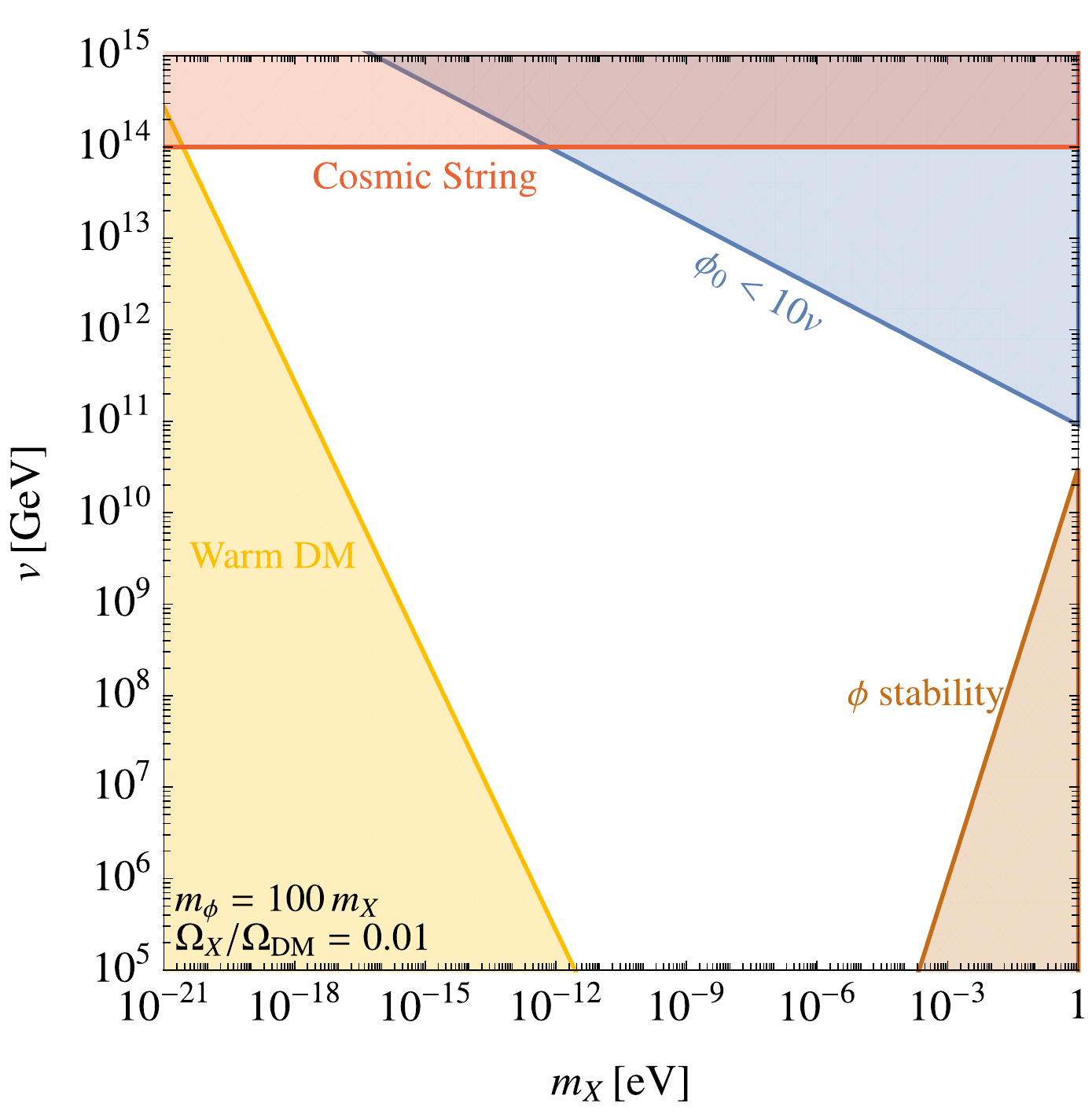}\hspace{0.5cm}
\includegraphics[width=8.25cm]{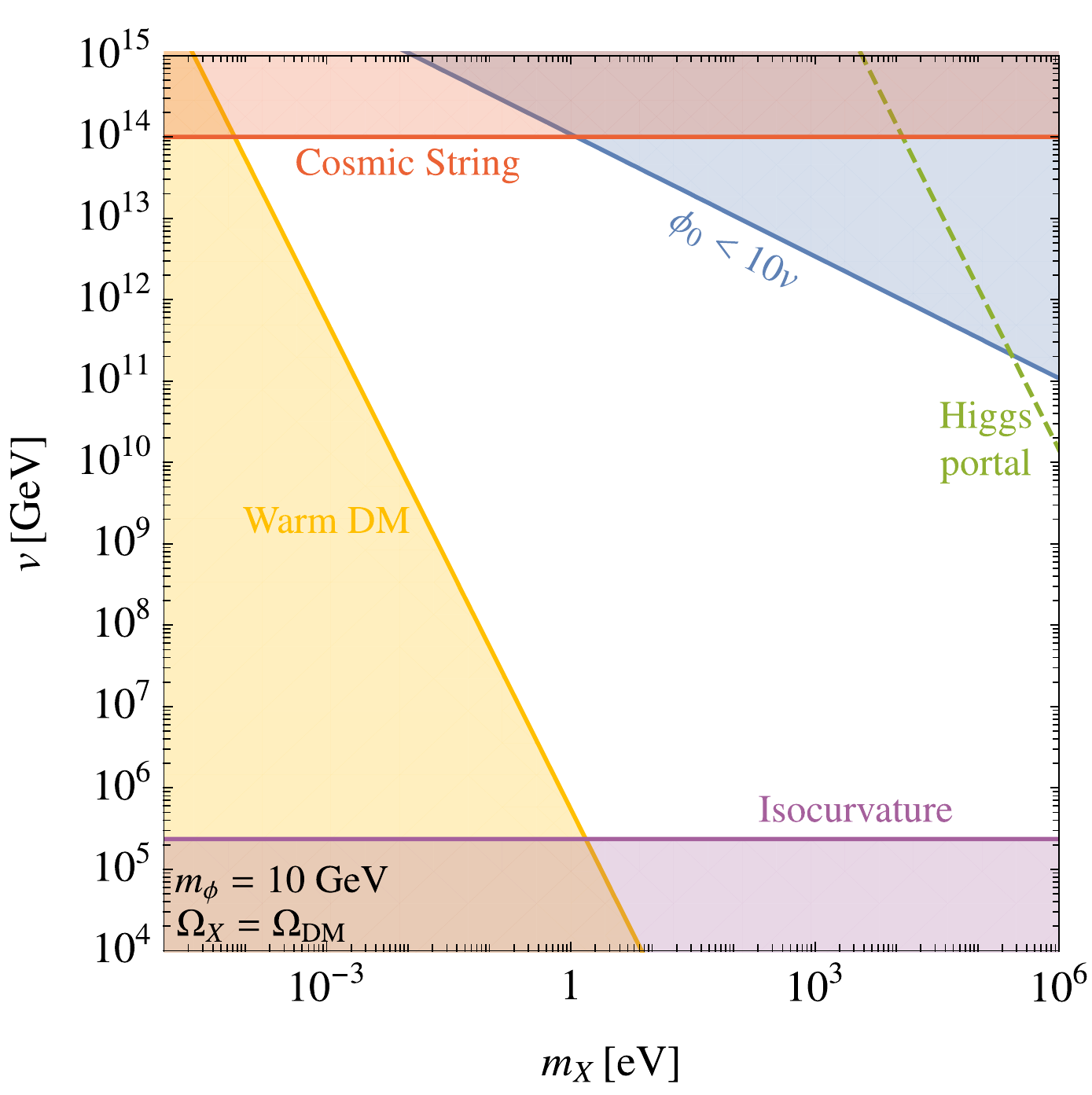}
\caption{Viable parameter space for parametric resonance production of vectors in the limit $ \lambda \gg e   $. 
We fix the initial condensate amplitude $ \phi _0 $ such that the dark sector saturates the DM relic abundance $\Omega_X + \Omega_\phi \simeq \Omega_\text{DM}$. 
Shown are constraints from coldness ({\bf \color{c8} yellow}), cosmic strings ({\bf \color{c4}red}), isocurvature ({\bf \color{c9} purple}), late time dark Higgs decays ({\bf \color{c6} brown}) and sufficiently long PR ({\bf \color{c1} blue}) as described in the text. 
{\bf Left}: At every point in parameter space we fix $m_\phi \simeq 100 m_X $. 
We do not incorporate any additional interactions, and the dark Higgs makes up nearly all the DM, i.e. $\Omega_X \simeq 10^{-2} \Omega_\phi$.
{\bf Right}: Vectors make up all of the DM, and the dark Higgs is eliminated at late times. 
At every point in parameter space we fix $ m_\phi = 10 ~{\rm GeV} $ and show the corresponding constraint from thermalization requirements ({\bf \color{c3} green}) as described in Appendix~\ref{sec:appendix}\,.}
\label{fig:largelambda}
\end{figure*}

Given the relative energy densities between the scalar and vector, it is most natural that the dark Higgs constitutes nearly all of the DM today, with the vector being a subdominant component. 
In order for this scenario to be consistent with observations we require the dark Higgs be both non-relativistic by around a keV and satisfy the relic density condition. 
Requiring the dark Higgs yield to be the right relic abundance fixes the required initial field value:
\begin{equation}
\label{eq:DMphi}
\frac{ \phi _0 }{ M_{\rm pl} } \simeq 10 \left( \frac{ \lambda T_{\rm eq} ^2  }{ m _\phi ^2 } \right) ^{1/3} \,.
\end{equation} 
The temperature at which the Higgs becomes non-relativistic is given by
\begin{equation}
T_\text{NR} \simeq v \left(  \frac{\lambda M_{\rm pl} }{\phi_0} \right) ^{1/2} \simeq 0.5 \left(  \frac{v^2 m_\phi^2}{T_{\rm eq} } \right) ^{1/3}\,.
\end{equation}
If the fraction of vector DM is greater than a few percent, it must also be sufficiently cold; otherwise, the relic vectors will be a hot DM subcomponent which is ruled out by the cosmic microwave background~\cite{Ade:2015xua}.
We show the viable parameter space for a vector subcomponent of DM in Fig.~\ref{fig:largelambda} (left), fixing the fraction of vector DM to be $10^{-2}$ (with the rest made up by the dark Higgs).

Even if vectors make up a small fraction of the DM abundance, they may still be detectable. 
Nevertheless, it is interesting to consider the possibility that vectors make up all of the DM due to some dynamics which eliminated the dark Higgs yield at later times. 
In particular, it is possible to introduce additional couplings to the model that drastically alter the expected relic abundance of dark Higgses, without affecting the abundance of vectors produced from PR. 
(Although we might have naively suspected that the large initial yield of $\phi$ particles could simply decay away to vectors through the perturbative process~\eqref{eq:invisdecay}, such population of vectors constitutes an $\OO(1)$ hot DM component.)
For now we will take it as a given that $Y_X$ reproduces the entire observed DM density. This fixes the initial field value:
\begin{equation}
\label{eq:DM1}
\frac{ \phi _0 }{ M_{\rm pl} } \simeq 10 \left( \frac{ \lambda T_{\rm eq} ^2  }{ m _X ^2 } \right) ^{1/3} \,.
\end{equation} 
The $X$ population becomes non-relativistic when the universe is at a temperature:
\begin{equation}
\label{eq:coldness1}
T_\text{NR} \simeq m _X \left( \frac{M_{\rm pl} }{\lambda \phi_0} \right) ^{1/2} \simeq 0.5 \left(  \frac{ m _X ^4}{ T_{\rm eq}  \lambda^2}  \right) ^{1/3}\,.
\end{equation}
As before, we require $T_\text{NR} \gtrsim \text{keV}$. Note that the initial field amplitude has a maximum value consistent with the vector DM abundance and coldness constraints:
\begin{equation}
\label{eq:maxphi0}
\phi_0 \simeq 10 M_{\rm pl}  \left(  \frac{T_{\rm eq} }{{T_\text{NR}}} \right) ^{1/2} \lesssim 2 \times 10^{17} ~\GeV\,.
\end{equation}
This makes the condition on oscillation during radiation-domination ($\phi_0 < M_{\rm pl} $) trivially satisfied. 

We now return to the elimination of the dark Higgs yield in the above scenario. 
If we assume the vector constitutes all of DM as per~\eqref{eq:DM1}, then to avoid the dark Higgs dominating the energy density of the universe at an intermediate time the yield should have been destroyed by the temperature $  \sim T_{\rm eq} m _\phi / m _X  $. 
This is not a constraint, though a necessary condition for the above formula to hold as they assume radiation domination throughout. 
If, on the other hand, the universe has gone through a period of dark Higgs domination that later gets dumped into the SM this could have profound implications on small scale structure~\cite{Gelmini:2008sh,Erickcek:2011us,Erickcek:2015jza} and changes the predicted relic abundances. 
If we assume this matter-dominated era lasts until the dark Higgs reheats the universe to a temperature $T_\text{R}$, the resulting entropy production dilutes the yield of relic vectors $ Y_{X} \sim  T_\text{R}/m _\phi  $.
A concrete example of the such a cosmology occurs if the dark Higgs is able to thermalize with the SM. 
This generically requires the dark Higgs to have a substantial coupling to the SM, and as a result the allowed mass range of $\phi$ will be subject to experimental constraints.
The simplest interaction of the dark Higgs with the SM is a Higgs-portal coupling. As we show in Appendix~\ref{sec:appendix}, this has severe constraints from star cooling and rare meson decays below around $ 5 ~{\rm GeV} $. We show the viable parameter space for vector DM production in Fig.~\ref{fig:largelambda} (right), assuming the large dark Higgs yield is eliminated at late times before dominating the energy density of the universe. Here we fix the dark Higgs mass to be $m_\phi = 10 ~{\rm GeV} $ and show the requirements on dark Higgs thermalization through the Higgs portal interaction, leaving a detailed examination of the necessary conditions to Appendix~\ref{sec:appendix}. We do not explicitly show the parameter space for vector DM production in the case of dark Higgs domination though we have checked the lower reach in $m _X $ is ultimately the same as that in the case of no entropy production.

\subsection{Case 2: $e\gg\lambda$}

We now turn to the limit where the gauge coupling is much larger than the quartic (and so $ m _X \gg m _\phi $). Due to the effects of post-scattering, the co-moving number densities of $ \phi $ and $ X $ at late again become comparable and are given by,
\begin{equation}
\label{eq:yields1}
Y_{X} \simeq Y_{\delta \phi} \simeq \frac{ \lambda }{ e}\,\frac{0.01 }{\lambda^{1/2}} \left(  \frac{\phi_0}{\Mpl} \right) ^{3/2}\,.
\end{equation}
This difference in mass of $ \phi $ and $ X $ leads to vectors dominating the energy density at late times, and the dark Higgs becomes a subdominant component with a fractional abundance $\lambda/e$. In addition, vectors are produced non-relativistically with typical co-moving momentum $k _\ast \lesssim  e \phi_0$, while the dark Higgses are dominantly produced from vector fluctuations with a similar spectrum and are thus highly relativistic.

The observed DM abundance is reproduced for the initial field amplitude of
\begin{equation}
\label{eq:DM2}
\frac{\phi_0}{M_{\rm pl} } \simeq 10 \left(  \frac{e^2 T_{\rm eq} ^2}{\lambda m _X ^2} \right) ^{1/3}\,.
\end{equation}
The temperature at which the vectors become non-relativistic is given by:
\begin{equation} 
\label{eq:tNRvector}
T _{ {\rm NR}} \simeq 2 \mX \l \frac{\Mpl \lambda}{\phi_0 e^2} \r^{1/2} \simeq 0.5 \l \frac{\mX^4 \lambda^2}{\Teq e^4}\r^{1/3}.
\end{equation} 
Note that by $T_\text{NR}$, the vector mass (initially dominated by fluctuations of $\phi$ after PR) assumes the vacuum value.  
Since the vector makes up most of the DM, we require the coldness constraint $ T _{ {\rm NR}} \gtrsim {\rm keV} $.
On the other hand, if the fraction of produced dark Higgses is roughly greater than $10^{-2}$, then this subdominant component must also be sufficiently cold. 

\begin{figure*}
\includegraphics[width=8.25cm]{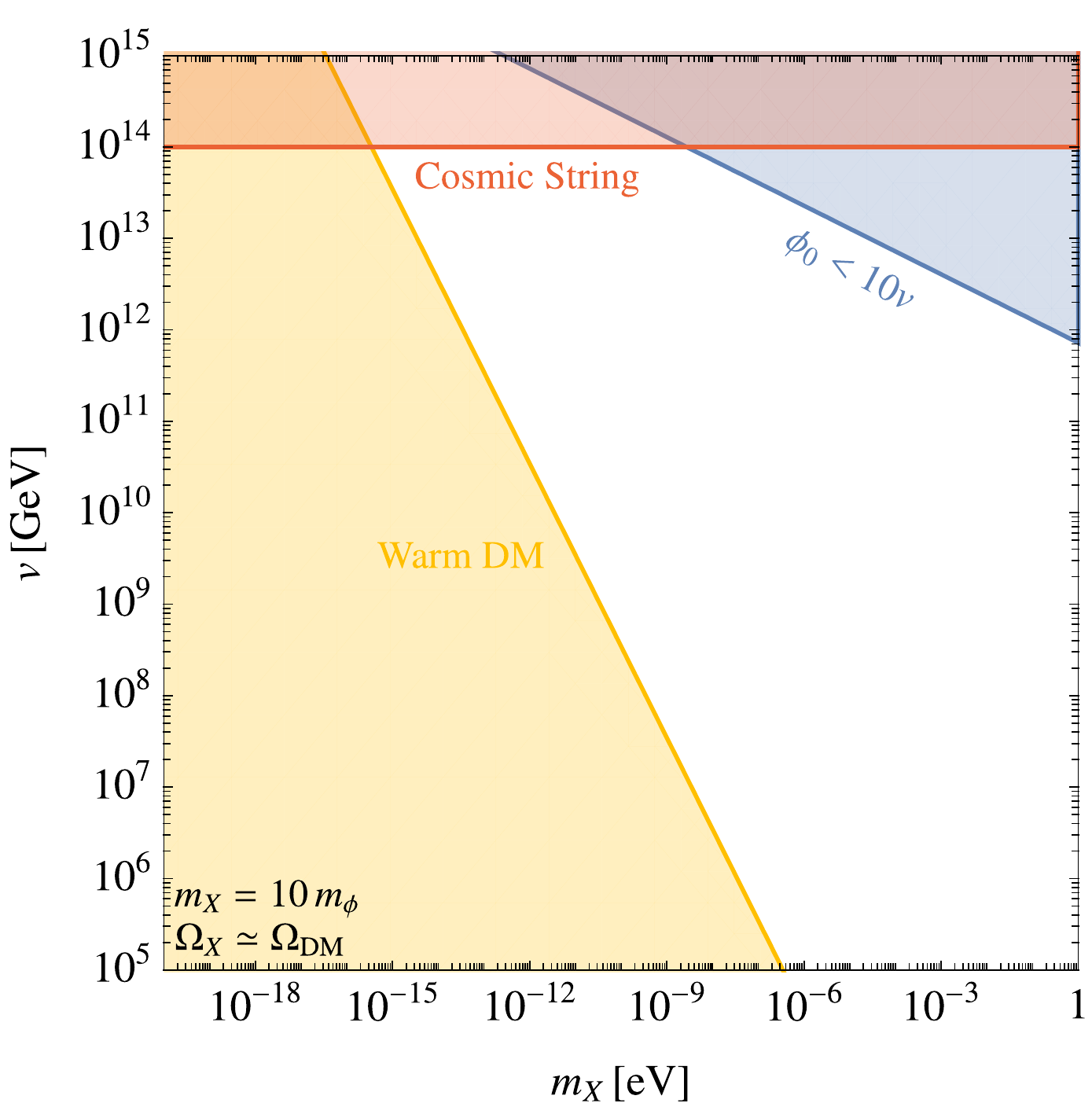}\hspace{0.5cm}
\includegraphics[width=8.25cm]{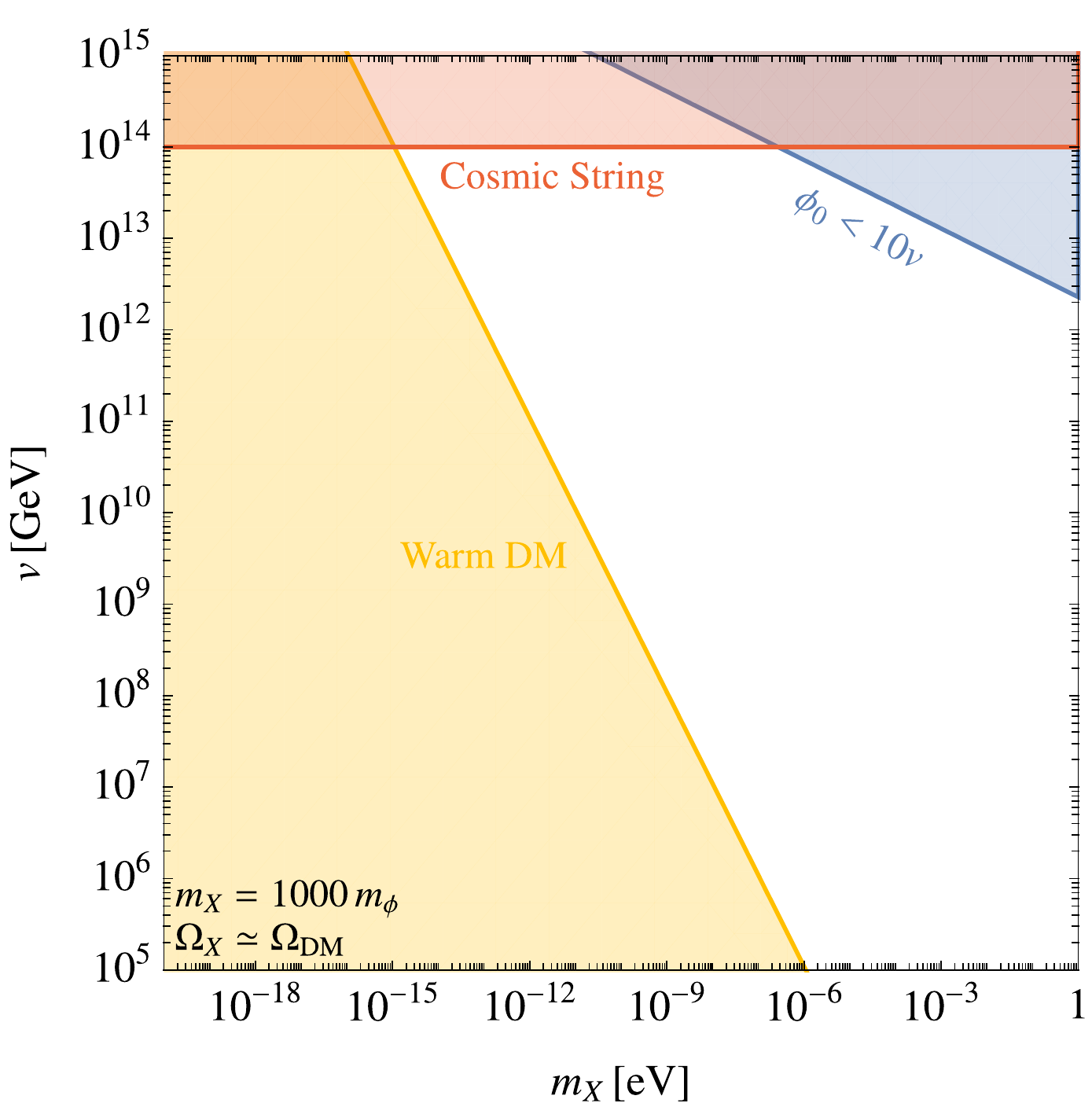}
\caption{Viable parameter space for parametric resonance production of vectors in the limit $ e \gg \lambda   $. 
We fix the initial condensate amplitude $ \phi _0 $ such that the dark sector saturates the DM relic abundance $\Omega_X + \Omega_\phi \simeq \Omega_\text{DM}$. 
Shown are constraints from coldness ({\bf \color{c8} yellow}), cosmic strings ({\bf \color{c4}red}), and sufficiently long PR ({\bf \color{c1} blue}) as described in the text. 
Both plots have the vector making up nearly all the DM. 
{\bf Left}: At every point in parameter space we fix $m_X \simeq 10 m_X$ and thus $\Omega_\phi \simeq 10^{-1} \Omega_X$. 
{\bf Right}: At every point in parameter space we fix $m_X \simeq 1000 m_\phi $ and thus $\Omega_\phi \simeq 10^{-3} \Omega_X$.}
\label{fig:smalllambda}
\end{figure*}

We show the viable parameter space for vector DM production in Fig.~\ref{fig:smalllambda} for $ e/\lambda = 10  $ (left) and $ e / \lambda = 10 ^{ 3} $ (right), with the value of $ \phi _0 $ fixed at every point to achieve the correct relic abundance. For $e/\lambda = 10$ the dark Higgs is a non-negligible subcomponent and in addition to the vectors being sufficiently cold we also require the dark Higgs is non-relativistic by a keV, while for $e/\lambda = 10^3$ we only require that the vector population satisfies the coldness constraint~\eqref{eq:tNRvector}. The lowest possible vector masses can be obtained by saturating $e \rightarrow \lambda $ where we find we can produce cold DM for $ m _X \gtrsim 10 ^{ - 18}~{\rm eV} $ (though saturating this limit results in the vectors being accompanied by non-negligible dark Higgs abundance).

\section{Phenomenology}
\label{sec:pheno}
In this section, we summarize some distinctive features of vector production through parametric resonance which could be used to differentiate it from other non-thermal cosmologies.

\paragraph{Dark Higgs}
Perhaps the most prominent prediction of PR production would be searching directly for the accompanying light scalar. The detectability of the scalar depends on its model-dependent coupling (if any) to the SM, and in general no such coupling is required to produce vectors. However, if $ \lambda \gg e $ than the scalar is either a large fraction of the DM abundance today, or the scalar is destroyed by some additional mechanism (e.g. thermalization with the SM) so that vectors make up all of the observed DM abundance. If the scalar is a non-negligible relic today then it could be searched for directly through experiments sensitive to light scalars. Furthermore, if it dominates the DM density then it could be observed as a (cosmologically slow) dark decay into the vectors from anomalous changes of equation of state of the universe~\cite{Gong:2008gi}. The current consistency with the $ \Lambda $CDM picture allows us to set a bound on this decay rate as given in Fig.~\ref{fig:largelambda}.

Alternatively, if the dark Higgs is assumed to thermalize with the SM then the minimal required coupling to achieve thermalization sets a convenient target for experimental searches. We study these specific requirements in the context of a Higgs portal coupling in Appendix~\ref{sec:appendix}.

\paragraph{Cosmic Strings}
As we have seen, PR produces large quantum fluctuations in the $X$ and $\phi$ fields. These fluctuations can lead to a large positive effective mass for $ \phi $ resulting in the symmetry being temporarily restored once PR terminates and a subsequent non-thermal phase transition once the mass of $\phi$ becomes negative~\cite{Kofman:1995fi}. This has an intriguing prediction of the formation of cosmic strings~\cite{Tkachev:1998dc}. Cosmic strings are one-dimensional topological defects, characterized by a string tension $\mu \sim v^2$. After formation, it is expected that the string network quickly approaches a scaling regime, i.e., energy density in strings scales with the energy density of the universe but roughly suppressed by the factor $G \mu$, where $ G = 1 / 8\pi M_{\rm pl} ^2  $ is Newton's constant. Such strings have several characteristic predictions owning to their induced large energy gradients in the universe. Perhaps the most robust detection of cosmic strings can be extracted from the cosmic microwave background, whose gravitational interaction would induce small temperature distortions leading to inhomogeneities in the temperature map~\cite{Kaiser:1984iv}. Using the WMAP data with a combination of cosmological observations such strings have yet to be observed, putting a constraint $G \mu \lesssim 10^{-7}$~\cite{Seljak:2006bg}. 

An additional prediction of cosmic strings comes from gravitational radiation emitted by the string oscillations. 
The evolution of a scaling cosmic string network is expected to contribute to the stochastic gravitational wave background~\cite{Siemens:2006yp} as well as induce gravitational wave bursts~\cite{Damour:2000wa}. This is contrast to global strings, which predominantly radiate massless Goldstone bosons (e.g., axion strings). The gravitational wave spectrum from a cosmic string network can be computed, under basic assumptions. The, thus far, null observation of a stochastic gravitational wave background by LIGO and pulsar timing arrays constrain $G \mu \lesssim 10^{-11}$~\cite{Ringeval:2017eww,Blanco-Pillado:2017rnf,Cui:2017ufi}, which roughly translates to a bound on the VEV $v \lesssim 10^{14} ~\GeV$. Future pulsar timing array measurements are expected to have improved sensitivity with the upcoming future Square Kilometer Array~\cite{Smits:2008cf} and provide an opportunity to probe these non-thermal phase transitions.

\paragraph{Isocurvature Perturbations}
Another prediction of this production mechanisms is due to the lightness of the dark Higgs, inducing isocurvature perturbations in the cosmic microwave background (CMB). During inflation, we presume $\phi$ is stuck with an initial field amplitude obeying $\lambda \phi_0 \lesssim H_\text{inf}$, and fluctuations, $ \delta \phi  \sim H_ { {\rm inf}} / 2\pi $. During PR the energy density of the $\phi$ condensate is transferred to the observed DM abundance and instills these isocurvature perturbations in the DM spectrum. These perturbations can be looked for in the CMB though they have yet to be seen~\cite{Ade:2015lrj}. This can be interpreted as a bound on the Hubble scale during inflation $H_\text{inf} \lesssim 3 \times 10^{-5} \phi_0$, which in the simplest picture suggests a bound $\lambda \lesssim 3 \times 10^{-5}$. This puts a relevant constraint for $ \lambda \gg e $ if the dark Higgs is required to thermalize with the SM but turns out to be negligible when we do not enforce this requirement. We note that, in principle, this isocurvature perturbation can be suppressed if the Hubble induced mass of $\phi$ is larger than $H_ { {\rm inf}}$.

\section{Discussion}
\label{sec:discussion}

In this work we present a new production mechanism for vector DM in the early universe through its (possible) coupling to a dark Higgs. 
The mechanism relies on the non-perturbative dynamics associated with parametric resonance, thus allowing the produced vectors to be ultralight while still being consistent with the stringent constraints on warm DM.

Vector production from parametric resonance has qualitative differences from the well-studied theory of scalar production. 
We study the equations governing the PR production of transverse and longitudinal modes and present an instability chart.
For $ \lambda \gg e $ the transverse mode production is highly inefficient while the longitudinal mode is rapidly produced (this can be understood as a consequence of the Goldstone boson equivalence theorem). Fluctuations of dark Higgses are also produced which results in a Higgs-dominated dark sector. In order for vectors to make up the entire DM abundance, additional interactions can be considered to thermalize the dark Higgs with the visible sector. We find produced vectors can be as light as $ 10 ^{ - 20 } ~{\rm eV} $ if they form $ 1 \% $ of the energy density in DM (with the dark Higgs making up the rest). If, on the other hand, we require the dark Higgs to thermalize with the SM it is difficult to foresee a viable model without making the dark Higgs heavier than around $ 10 ~{\rm GeV}  $ (otherwise there are tight constraints on its coupling). This restricts the produced vector DM to having masses above around $ 10 ^{ - 4}~ {\rm eV} $. 
In the case where $ e \gg \lambda $ both the transverse and longitudinal mode can be efficiently produced, though as a consequence of initial conditions set by inflation we still expect the longitudinal mode to dominate for a wide range of parameters. As in the previous case fluctuations of $ \phi $ are rapidly produced resulting in comparable number densities between vectors and dark Higgs, although due to the ratio of masses the DM energy density today is dominated by vectors. Ultimately, the coldness constraint restricts the viable vector DM mass to be above $10 ^{ - 18}~{\rm eV} $.

Our study of PR production of ultralight vectors was not meant to be exhaustive, and we conclude by commenting on directions we feel merit further attention. 
Firstly, the focus of this work was entirely on vectors which get their mass from a dark Higgs. In principle, this could easily be generalized to other types of scalars which obtain a large field value. Secondly, in this work we did not attempt a complete lattice simulation of the non-linear effects. This would be particularly important in the limit of $ e \gg \lambda $ since in this case it is possible that the coldness constraint is significantly weakened if the vectors do not get boosted to their maximum possible momenta, $ e \phi _0 $.
Furthermore, it is important to note that a general feature of this framework is the necessity for tiny couplings (for the mass range in the $ e \gg \lambda $ case, gauge couplings in the viable parameter space go down to as low as $ \sim 10 ^{ - 40} $). 
While such couplings are technically natural, it would be interesting to see how viable these are in a UV model.
Lastly, in this work we briefly explored the prominent phenomenological signatures of this production mechanism though it may be fruitful to consider these in more detail as well as others to differentiate PR production from other possible production mechanisms.

\section*{Acknowledgments}
We would like to thank Ryan Janish, Simon Knapen, Takemichi Okui and Surjeet Rajendran for useful discussions. 
This work is supported in part by the DOE under contracts DE-AC02-05CH11231 (JD) and DE-SC0009988 (KH).
%JD is supported in part by the DOE under contract DE-AC02-05CH11231.
\vspace{0.2cm}

{\em Note Added:} 
During the preparation of this work, we became aware of~\cite{Co:2018lka},~\cite{Agrawal:2018vin}, and~\cite{Bastero-Gil:2018uel}---these all discuss possible production of light vector DM from an oscillating axion field via the tachyonic instability.

\appendix
\section{Dark Higgs Thermalization}
\label{sec:appendix}
In this section we further examine vector production by PR for the case $\lambda \gg e$ requiring that the vector constitutes all of the DM today. In Sec.~\ref{sec:vectorDM}, we explored the parameter space in this scenario assuming that the large initial yield of dark Higgs after PR thermalizes with the SM. Here, we explicitly study the case of thermalization through a Higgs-portal coupling,
\begin{equation}
\label{eq:kappa}
{\cal L}  \supset y^2 |\varphi|^2 |H|^2\,,
\end{equation} 
where $H$ is the SM Higgs doublet. After electroweak symmetry breaking, there is a mixing between the two real scalars with an angle, 
\begin{equation}
\tan 2\theta \simeq y^2\frac{2  v v_{{\rm EW}}}{m_h^2 - m _\phi ^2}\,,
\end{equation}
where $v_{{\rm EW}} \simeq  246 ~{\rm GeV} $.  
For simplicity, we will consider the regime $m_\phi \ll m_h$ so that $\theta \sim  y^2 v/v_\text{EW}$. This interaction must satisfy a number of conditions in order for the cosmology to be viable, namely:
\begin{enumerate} 
\item  The thermalization occurs sufficiently rapidly,
\item  The dynamics of PR is largely unaffected,
\item  The coupling is not ruled out by experiments.
\end{enumerate}
Our aim here is to show the existence of a viable parameter region in dark Higgs mass and coupling for which all conditions can be consistently satisfied.

We first address the requirements on thermalizing the dark Higgs. 
The relevant processes differ if thermalization occurs before or after the electroweak phase transition which depletes the SM Higgs. 
Above this scale the dominant number-density depleting process that brings the dark Higgs into chemical equilibrium is through a Higgs absorption $\phi H \to H$. 
Proper calculation of this rate requires non-equilibrium field theory techniques. 
The thermalization rate is roughly of order~\cite{Mukaida:2012qn} $ \Gamma _{ \phi H \rightarrow H } \sim  y^4 v^2 / T $.
From this we estimate the temperature of dark Higgs thermalization $T_\text{th} \sim (y^4 v^2 \Mpl)^{1/3}$.
Requiring this thermalization temperature to be above $ \sim 100 ~{\rm GeV} $ puts a constraint on the mixing angle, $  \theta \gtrsim  10 ^{ - 7 } $. 
Below the electroweak scale, $\phi$ will continue to interact with SM fermions in the thermal bath.
For instance, thermal $\phi$ particles can scatter off fermions (e.g. quarks) in the plasma with a rate~\cite{Mukaida:2012qn}, $\Gamma_{\phi q \to q g} \sim \theta^2 y_f ^2 T$, where $y_f$ is the largest fermion Yukawa coupling still in the SM bath. For this to be above the mass of the fermion requires $ \theta \gtrsim 10 ^{ - 8 } / \sqrt{ y _f } $. This ensures $\phi$ has a thermal abundance and thus dumps its energy to the SM bath when $ T \lesssim m _\phi $. 
If, e.g., $m_\phi = 10~\GeV$, this process is in thermal equilibrium with bottom quarks by the time the temperature drops to $T \sim m_b$ as long as $\theta \gtrsim 10^{-7}$.  

Next we consider the implications the Higgs portal coupling on the PR mechanism. 
As long as the coupling is not fine-tuned $\theta<m_\phi/m_h$, the SM Higgs mass correction at the time of PR is less than the expected thermal mass. 
Therefore, the Higgs portal coupling gives a mass correction $\delta m _\phi ^2 \sim y^2 T_\text{osc}^2$ to the dark Higgs at the onset of oscillations. 
If this is larger than the assumed effective mass-squared $m_\text{eff}^2 \sim \lambda^2 \phi_0^2$, the dark Higgs would have oscillated earlier with a large frequency (set by the size of $ y ^2 $), thus rendering the PR production of vector modes narrow and relatively inefficient. 
Requiring this thermal contribution to be sufficiently small $\delta m_\phi^2 \lesssim  \lambda^2 \phi_0^2$ translates into an upper bound on the mixing angle,
\begin{align}
\label{eq:thpotential}
\theta &\lesssim  30 \times \left( \frac{m_\phi}{v_{\rm{EW}}} \right) \left(  \frac{\phi_0}{M_{\rm pl} } \right) \,,\\
& \simeq  10^{-2} ~\left( \frac{m _\phi }{\GeV} \right) \left( \frac{\phi_0}{ 10 ^{ - 1} M_{\rm pl}} \right)\,. \nonumber
\end{align}
For Fig.~\ref{fig:largelambda} where we fix $ m _\phi \simeq 10 ~{\rm GeV} $, this is the dominant upper bound on the mixing angle.

A second possible effect of the Higgs portal coupling is PR production of the SM Higgs. 
For thermalization we generically require $y \gg \lambda$ which would suggest that PR may be efficient in producing the SM Higgs; this is ultimately not the case due to the large thermal mass of the Higgs. 
We can verify this by computing the adiabatic parameter $R = |\dot{\omega}|/\omega^2$ for the SM Higgs with a time-dependent frequency $\omega(t) \simeq \sqrt{k^2 + m _H ( T ) ^2  + \frac{1}{2} y^2 \phi(t)^2}$. 
Since the thermal mass is strictly larger than the mass contribution from the oscillating field at the time of PR ($y^2 \phi_0^2 \ll  T_\text{osc}^2$), the adiabatic parameter is always less than unity. 
We conclude that there is no significant non-perturbative production of the SM Higgs. 

There are a number of phenomenological constraints on a Higgs-portal scalar. 
These constraints depend on whether the scalar decays visibly or appears invisible (at least on detector scales.) For detailed review see, e.g.,~\cite{Krnjaic:2015mbs}. For low masses $m_\phi \lesssim  {\rm MeV} $, we find stellar cooling constraints~\cite{Hardy:2016kme} are in tension with the requirements of thermalization. For intermediate masses, $ {\rm MeV} \lesssim  m _\phi \lesssim  300~{\rm MeV} $ constraints from Supernova 1987A and rare Kaon decays are powerful, but it may be possible to have a scalar in this mass range with a sufficiently large $\theta$ consistent with thermalization if it sits in the small gap between these constraints.
For masses $\GeV \lesssim m _\phi  \lesssim 5 ~\GeV$, rare B meson decays roughly constrain $\theta \lesssim 10^{-3}$.
Above this scale, scalar production at LEP constrains $\theta \lesssim 10^{-1}$ although this is weaker than the condition~\eqref{eq:thpotential}.  
Interestingly, if $\phi$ decays visibly then lower values of $\theta$ in this mass range could also be probed by future experiments designed to look for long-lived particles~\cite{Alekhin:2015byh,Chou:2016lxi,Gligorov:2017nwh,Feng:2017uoz,Gligorov:2018vkc}.

Lastly one may wonder if the dark Higgs ever dominates the energy density of the universe.
Dark Higgs domination will take place if the temperature of thermalization is less than $ \lesssim  T_{\rm eq}  (m_\phi/\mX)$, or in terms of the mixing angle:
\begin{align}
\label{eq:thbeatsdom}
\theta & \lesssim  \left( \frac{\Teq^3}{v_{\rm{EW}}^2 M_{\rm pl} } \right) ^{1/2} \left( \frac{m _\phi }{\mX} \right) ^{3/2} \,,\\
&\simeq  5 \times 10^{-6} \left( \frac{m _\phi }{\GeV} \right) ^{3/2} \left( \frac{10^{-4}~ {\rm eV} }{m _X } \right) ^{3/2} \,. \nonumber
\end{align}
We see that for $ m _\phi \simeq 10 ~{\rm GeV} $ and the relevant vector mass range $m _X  \gtrsim 10^{-4}$ in Fig.~\ref{fig:largelambda}, this is never the case as long as we satisfy the condition of thermalization. 

\bibliography{dpdm}
\end{document}